# Superconducting MgB$_2$ Thin Films with $T_c \approx$ 39K Grown by Pulsed Laser Deposition

S.F.Wang, S.Y. Dai, Y..L.Zhou, Z.H.Chen, D.F.Cui, J.D.Yu, M.He, H.B.Lu, and G.Z.Yang

*Laboratory of Optical Physics, Institute of Physics and Center for Condensed Mater Physcis, Chinese Academy of Sciences,*

*Beijing 100080, Peapoles Repulic of China*

**Superconducting MgB$_2$ thin films were fabricated on Al$_2$O$_3$(0001) substrates under an ex-situ processing conditions. Boron thin films were deposited by pulsed laser deposition followed by a post-annealing process. Resistance measurements of the deposited MgB$_2$ films show $T_c$ of ~ 39K, while scanning electron microscopy and x-ray diffraction analysis indicate that the films consist of well-crystallized grains with a highly c-axis-oriented structure.**

The discovery of superconductivity in MgB$_2$ with $T_c \approx$ 39K by Akimistu and coworkers[1,2] offers a new class of simple, low-cost and high-performance binary intermetallic superconductors with a record high superconducting transition temperature for a nonoxide and non-*C60*-based compound. A significant experimental effort has been focused on the growth of MgB$_2$ films[3-7] for the availability of such a film can be of great value for further basic studies and electronic applications. However, the growth of MgB$_2$ is complicated due to the large differences in vapor pressure between boron and magnesium and the high sensitivity of magnesium to oxidation. To overcome this two complicated factors for the fabrication of MgB$_2$ thin films, an ex-situ method is exploited, which are based on a similar approach as used in MgB$_2$ wires:[8] magnesium atoms diffuse into boron films at high temperature and high magnesium pressure.

In this paper, we report our results on the growth of $MgB_2$ thin films by an ex-situ method. First, boron films were deposited on $Al_2O_3$(0001) substrates at 900℃ under the back ground pressure of $6\times10^{-4}$Pa by pulsed laser deposition. During deposition, an XeCl excimer-laser beam at a wavelength of 308nm was focused on the surface of the boron target. The laser spot sizes on the target were about 1.5mm$^2$ and the energy density at the target surface was 10J/cm$^2$. The laser was operated at 6Hz. Approximately 350--400nm thick boron films were prepared with a deposition rate of about 1Å/s. After deposition, the boron films were dipped into the high pure alcohol for 30min to get rid of the $B_2O_3$. Then the films together with magnesium and $MgB_2$ pellets, which were wrapped with tantalum foil, were annealed in an evacuated quartz tube at 900℃ for 60min, after that, the samples were allowed to cool slowly to room temperature. The resistance measurement was performed with a standard four-probe technique. Scanning electron microscopy (SEM) and x-ray diffraction (XRD) were used to studied the properties of the $MgB_2$ films.

Figure 1 shows the temperature dependence of the resistance of $MgB_2$ thin films under zero magnetic field, the onset and zero-resistance temperatures are about 39K and 36K, respectively. A very sharp transition with a width of about 0.8K from 90% to 10% of the normal state resistance is observed, which is comparable to most of the high-quality bulk $MgB_2$ samples.

*Fig.1 Resistance versus temperature for $MgB_2$ thin films on $Al_2O_3$(0001) substrates.*

The surface morphology of the $MgB_2$ films grown on $Al_2O_3$(0001) substrates is shown in Fig.2, well-defined, large hexagonal grains(1-2μm) and a porous appearance are observed in the magnified view of Fig. 2. X-ray energy dispersive spectroscopy (EDX) analysis shows only a small amount of oxygen contamination(0.3%) in the deposited $MgB_2$ films.

*Fig.2   SEM image of the $MgB_2$ thin film with $T_c \sim 39K$ on $Al_2O_3(0001)$ substrates. The left one is a magnified view.*

Figure 3 presents a typical x-ray diffraction pattern for the $MgB_2$ films grown on $Al_2O_3(0001)$ substrates. Two sharp peaks corresponding to the $MgB_2$ (001) and (002) line indicates that most grains in the film have their *c*-axis orientation, the other two small peaks correspond to the remnant of Mg(101) and MgO(222) lines, which may be a possible explanation for lower transition temperature of $MgB_2$ thin films when compared to its bulk materials.

*Fig.3   XRD patterns of the $MgB_2$ thin film with $T_c \sim 39K$ on $Al_2O_3(0001)$ substrates.*

In summary, thin films of $MgB_2$ with $T_c \sim 39K$ have been fabricated on $Al_2O_3(0001)$ substrates. The deposition method was a two-step ex-situ approach. A deposition of boron films by pulsed laser deposition followed by a high magnesium pressure anneal step at 900℃. The technique results a small amount of oxygen contamination, well-crystallized $MgB_2$ thin films with a highly *c*-axis-oriented structure.

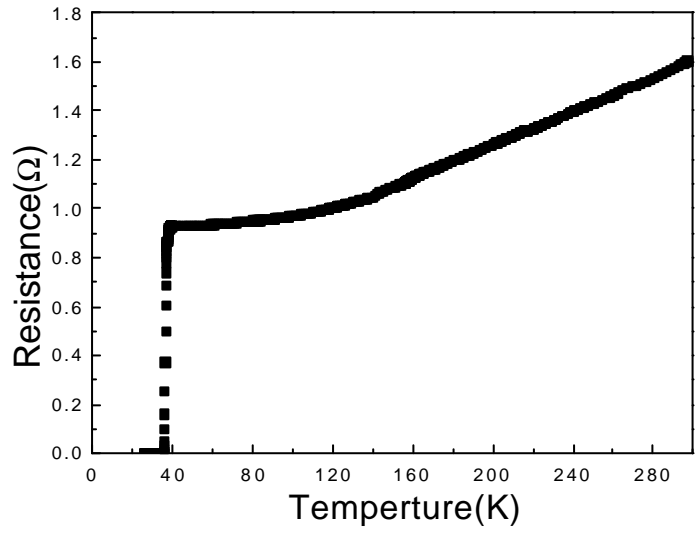

Fig.1   Resistance versus temperature for MgB$_2$ thin films on Al$_2$O$_3$(0001) substrates.

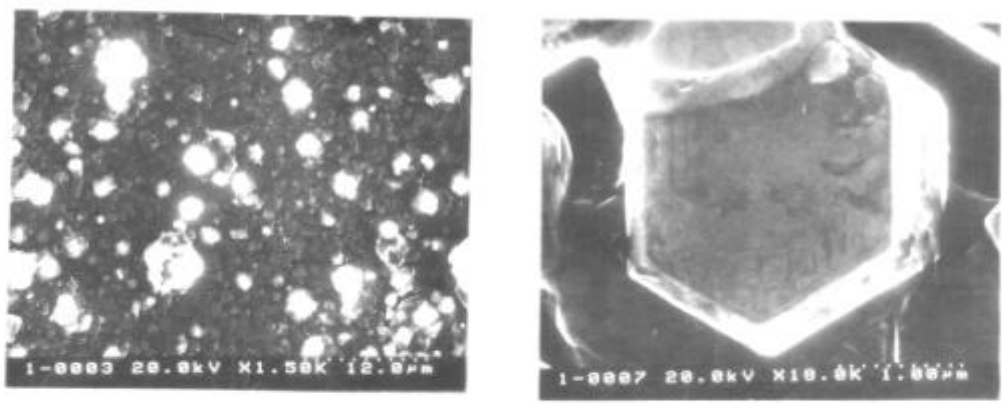

Fig.2 SEM image of the MgB$_2$ thin films with T$_c$~39k on Al$_2$O$_3$(0001) substrates. The left one is a magnified view.

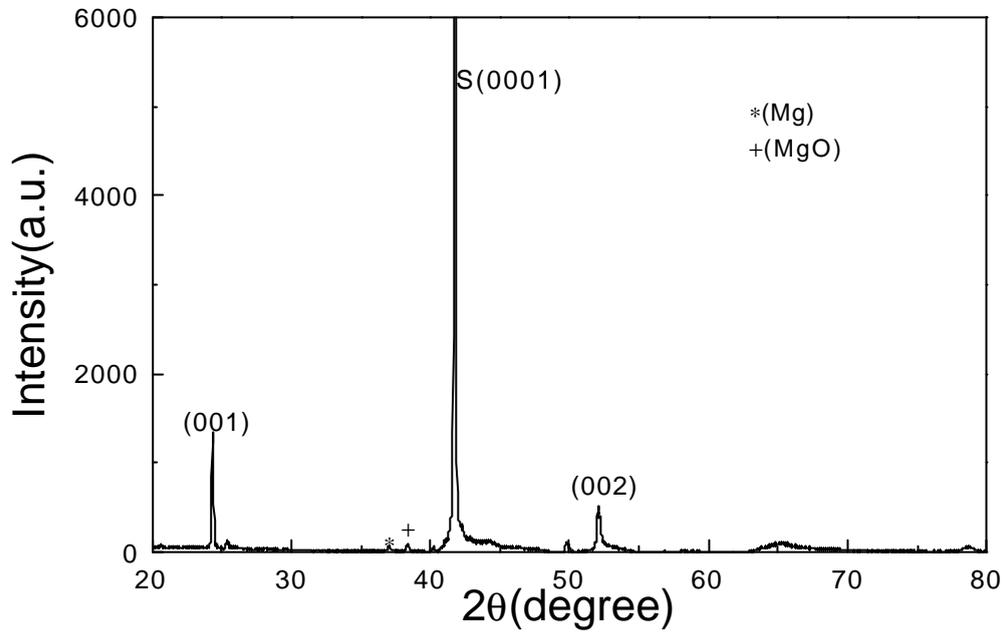

Fig.3 XRD patterns of the MgB$_2$ thin film with T$_c$~39k on Al$_2$O$_3$(0001) substrates.